\documentstyle[aps,preprint]{revtex}

\begin{document}

\draft

\title{
 Hadronic Light-by-light Scattering Effect
 on Muon $ g - 2 $ }

\author{
 M. Hayakawa$ ^1$ \footnotemark[1],
 T. Kinoshita$ ^2$ \footnotemark[2],
 and A. I. Sanda$ ^1$ \footnotemark[3] }
\footnotetext[1]
{
 Electronic address: hayakawa@eken.phys.nagoya-u.ac.jp
}
\footnotetext[2]
{
 Electronic address: tk@hepth.cornell.edu
}
\footnotetext[3]
{
 Electronic address: sanda@eken.phys.nagoya-u.ac.jp
}

\address{
 $ ^1 $  Department of Physics, Nagoya University,
         Nagoya 464-01 Japan \\
 $ ^2 $  Newman Laboratory, Cornell University,
         Ithaca, New York 14853 USA }

\preprint{DPNU-94-21}

\date{\today}

\maketitle

\begin{abstract}
 The hadronic light-by-light scattering
contribution to
muon $g-2$ is examined using low energy effective theories of QCD,
the Nambu-Jona-Lasinio model
and  hidden local chiral symmetry, as guides.
 Our result is $- 36 \times 10^{-11}$
with an uncertainty of $\pm 16 \times 10^{-11}$,
which includes our best estimate of model dependence.
 This is within the expected measurement uncertainty
of $40\times 10^{-11}$
in the forthcoming experiment at Brookhaven National Laboratory.
 Our result removes one of the main theoretical obstacles
in verifying the existence of
the weak contribution to the muon $g-2$.
\end{abstract}

\pacs{ PACS numbers: 11.15.Tk, 11.10.Rd, 11.40.Ha,
       13.10.+q, 14.60.Ef }

\narrowtext

 The anomalous magnetic moment
of the muon $ a_\mu \equiv \frac{1}{2}
\left( g_\mu - 2 \right) $ is one of the basic physical observable
that can be measured with high precision.
 The most recent results of
experiment and theory are as follows \cite{CERN,KM}:
\begin{eqnarray}
 a_\mu ({\rm exp}) &=& 1\ 165\ 923\ (8.5) \times 10^{-9} ,
  \nonumber \\
 a_\mu ({\rm th}) &=& 116\ 591\ 877 (176) \times 10^{-11}.
 \label{amu}
\end{eqnarray}
 The uncertainty in the measurement of $a_\mu ({\rm exp})$
will be reduced in the new experiment
at the Brookhaven National Laboratory to
$
 \sim 40\times 10^{-11}
$
\cite{hughes}.
 This is about one-fifth of
the one-loop weak correction \cite{weak}
\begin{equation}
 a_\mu({\rm weak}\mbox{-}1)=195\ (1) \times 10^{-11},
\end{equation}
and of the same order of magnitude
as the leading logarithmic term
of the two-loop electroweak correction
\cite{weak2},
$
a_\mu({\rm weak}\mbox{-}2)=-42 \times 10^{-11},
$
offering an exciting opportunity to
test the quantum effect of the electroweak theory.

\par
 Before comparing theory with the forthcoming measurement,
however, it is necessary to reduce further
the uncertainties
in the theoretical prediction for the hadronic contribution.
 The largest uncertainty comes
from the hadronic vacuum-polarization contribution,
$ a_\mu({\rm had.v.p.}) $
\cite{Bouchiat}.
 Fortunately,
this contribution can be expressed as a convolution of
known function
with the experimentally measurable quantity $R$,
the ratio of the hadron production cross section to
the $ \mu^+ \mu^- $ production cross section
in $e^+ e^-$ collisions.
Recent measurements of $R$
at VEPP-2M will improve this estimate significantly \cite{worstell}.
Together with future measurements at DA$\Phi$NE, BEPC, etc.,
the error in $a_\mu$(had.v.p.) will be reduced
to the level of the upcoming experimental limit.

 The contribution of the hadronic light-by-light
scattering diagram shown in Fig. 1
is potentially a source of more serious difficulty
because it cannot be expressed in terms of
experimentally accessible observables
and hence must be evaluated by purely theoretical consideration.
Recently, some doubts have been raised
\cite{Barbieri,Einhorn}
about the reliability of previous estimate \cite{K-had}.
 In view of its importance in interpreting
the experiment and
in drawing inferences about potential $``$new physics'',
we have decided to reexamine its theoretical basis.

\par
 The bulk of
hadronic light-by-light scattering contribution
to $a_\mu$ is determined by the dynamics around the muon mass
$ m_\mu \simeq 106\,{\rm  MeV} $,
which is right in the midst of the non-perturbative regime of QCD.
 What we need is a reliable evaluation of
an off-shell four-point function at these energies.
 In view of the current status of QCD, however,
it is not an easy job to carry out such a calculation
from first principles.

 Fortunately, this energy region is populated mostly by pions,
and a considerable
information is available about low energy pion dynamics.
 Chiral symmetry governs most of it.
 However,  higher energy regions may also contribute
significantly to $a_\mu$.
 Momentum-expanded amplitudes obtained
in a systematic chiral expansion,
however, cannot be introduced directly
into Feynman graphs for muon anomaly since it
leads to divergent integrals.
 To get around
this difficulty, Ref. \cite{deRafael} used
the Nambu-Jona-Lasinio (NJL) model.
 We chose to rely on the
hidden local symmetry (HLS) approach \cite{Bando}.
 Since the HLS Lagrangian
can be derived from the extended NJL ( ENJL ) model
\cite{ebert},
these two
approaches are equivalent
as far as application to our problem is concerned.


(a) Relevant diagrams:

 The NJL model and the $1/N_c$ expansion
suggest three distinct contributions
to the light-by-light scattering amplitude
at low energies \cite{deRafael}.
 Their contributions to $a_\mu$ are shown
in Fig. 2.
 Fig. 2(a)
shows the charged pseudoscalar meson loop contribution.
 It is $ {\cal O}(1) $ in  $1/N_c$ expansion
and $ {\cal O}(p^4) $ in chiral perturbation.
  Fig. 2(b) shows
one pion pole diagram which is ${\cal O}(N_c)$
and $ {\cal O}(p^6) $.
  Fig. 2(c) shows
a quark loop diagram which is ${\cal O}(N_c)$
and starts at $ {\cal O}(p^8) $ in chiral expansion.

 From the viewpoint of QCD,
the single-quark-loop diagram Fig, 2(c)
represents the averaged hadronic continuum effect
in a certain energy region.
At low energies,
there should be higher-order QCD corrections, which
can be approximated by a pion loop diagram
 Fig. 2(a)
and the Nambu-Goldstone boson pole diagram
Fig. 2(b).
 Note that the latter two exclude
a single-quark-loop contribution
since the pion loop requires at least two quark loops
and the pion pole starts from a diagram
in which at least two gluons propagate
between a quark and an antiquark forming the pion.

 Contributions involving more loops
will be suppressed by a factor ${m_\mu}/{(4\pi f_\pi)} \sim 1/11$.
 Diagrams
in which the pion loop
 in Fig. 2(a), $\pi^0$ pole in
Fig. 2(b), and $u,d$ quarks
in Fig. 2(c)
are replaced by kaon loop, $\eta$ pole,
and $s$ quark, respectively,
may also be nonnegligible.


(b) Imbedding the light-by-light scattering amplitude:

 When the light-by-light scattering amplitudes
are included in the Feynman graphs for $g-2$,
photons must be taken off-shell.
 The validity of low energy approximation to these amplitudes
will then be affected
by the convergence of photon momentum integration.

 In order to see what kind of problems might be in store,
let us begin by comparing two previous treatments of
the contribution to $a_\mu$
due to the charged pion loop diagram
of Fig. 2(a)
\cite{K-had}.
 One deals with the point-like
$\gamma\pi\pi$ and
$\gamma\gamma\pi\pi$ couplings (namely the scalar QED),
which yields
\begin{equation}
  a_\mu({\rm sQED})
 = -0.035\ 57~(18) \left( \frac{\alpha}{\pi} \right)^3 .
 \label{sQED}
\end{equation}
(This is a slight improvement
in statistics  over that of Ref. \cite{K-had}).
 The second approach attempts to improve (\ref{sQED})
based on the vector meson dominance model in which photons
couple to charged pions through $\rho$ meson.
 In Ref. \cite{K-had},
this was achieved by modifying the photon propagator as
\begin{equation}
 \frac{i}{q^2} \rightarrow \frac{i}{q^2}
 \frac{m^2_\rho}{m^2_\rho - q^2}=
 \frac{i}{q^2} - \frac{i}{q^2 - m^2_\rho} ~.
 \label{sut}
\end{equation}
 This leads to
\begin{equation}
  a_\mu({\rm nVMD})
 = -0.01\ 25(19) \left( \frac{\alpha}{\pi} \right)^3 .
 \label{KNO-VMD}
\end{equation}
 Comparing (\ref{sQED}) and (\ref{KNO-VMD})
we see that the introduction of the $\rho$ meson
makes a big difference.
 This looks alarming.
 After all, vector mesons hardly contribute
to  $\pi\pi$
scattering near threshold since the chiral symmetry demands
that their contribution vanishes in the soft pion limit.
 Why should they make such a difference, then ?
 Actually, this question is not posed correctly since the role
of the $\rho$ meson in (\ref{KNO-VMD})
is primarily that of modifying
the photon propagator
and not directly related to the $\pi$-$\pi$ interaction.
 What is more important is to check whether
the vector meson contribution
is properly included in (\ref{KNO-VMD}).

 It was pointed out in Ref.
\cite{Einhorn}
that the  prescription given by (\ref{sut})
does not respect the Ward identities
for the coupling of photons to charged pions,
even though it maintains gauge invariance.
 Upon  closer examination,
we found further that the substitution (\ref{sut})
is not consistent with chiral symmetry.

 In this paper, we solved these problems
based on the HLS approach \cite{Bando}.
 This is a convenient way to introduce
the dynamics of pions and vector mesons
preserving chiral symmetry and gauge invariance.
 It should be noted that this approach reproduces
all current algebra results, such as the KFSR relation,
as low energy theorems.
 In this sense, it is the leading candidate for the extension
of chiral dynamics of pions
to include vector mesons.
 Actually, insofar as only the low
energy dynamics is relevant for the computation of $a_\mu$,
any model consistent with chiral symmetry
should yield a similar contribution to $a_\mu$.

Before presenting our result, some comments relevant for
computing  $a_\mu$ are in order.

1.  The most important feature of the HLS Lagrangian,
in computing $a_\mu(a)$ which corresponds to Fig. 2(a),
is that it
does not have the $\rho^0\rho^0\pi^+\pi^-$
coupling.
 The naive substitution (\ref{sut})
assumes the presence of this coupling in the Lagrangian.
 Indeed we found this to be the source of the problem
pointed out in \cite{Einhorn}.
Ward-Takahashi identity is satisfied once
this is corrected.

2.  In computing $a_\mu(b)$,  the naive VMD model adopted in
Ref. \cite{K-had} did not have strong
theoretical basis beyond
that it provided an effective UV cut-off.
 It has been shown recently, however,
that it is justifiable within the HLS approach
\cite{Harada},
at least as far as going off-shell with respect to
photon momenta is concerned.
 This is also realized in the ENJL model,
 in which the $ \pi^0 $ pole diagram
contains two triangle loops of constituent quarks
and
$ \rho $ meson is allowed to propagate
before the quark couples to
the photon.

 3.  To compute $a_\mu(c)$,
it is necessary to know how the quark couples to the photon
and how vector mesons come into the picture.
 In this respect we are guided again by
the ENJL model in which a quark loop couples to a photon through
a vector meson (see Fig. 4 of Ref. \cite{deRafael} ).


(c) Numerical results:


 An extensive numerical evaluation of
the contribution of Fig. 2
to $a_\mu$,
within the framework of HLS, has yielded
\begin{eqnarray}
  a_\mu(a)
 &=& - 0.003~55~(12) \left( \frac{\alpha}{\pi} \right)^3 ,
 \nonumber   \\
 a_\mu(b) &=&
 -0.026\ 94(5)
 \left( \frac{\alpha}{\pi} \right)^3,
 \nonumber   \\
 a_\mu(c)
 &=&~ 0.007\ 72(31) \left( \frac{\alpha}{\pi} \right)^3.
 \label{qloop}
 \end{eqnarray}
 The errors quoted above are
those of numerical integration only
and do not include estimates of model dependence.
 The result $a_\mu (a)$
is much smaller than (\ref{sQED}).
 Similarly,
$a_\mu (c)$ is
considerably smaller
than the corresponding results without vector mesons.
$a_\mu (b)$ has a sign opposite to that of
Ref.\cite{K-had}, which had a sign error in some terms.
 Thus,
the difference between (\ref{sQED}) and (\ref{KNO-VMD}),
which worried us a great deal, does not go away
in spite of the improved theory which preserves
chiral symmetry and Ward-Takahashi identity.
 This forced us to examine
which regions of momentum space dominate in $a_\mu$.
We explored this problem by varying masses
instead of examining momentum dependence directly.

(d) Mass dependence:

 The dependence of $a_\mu$
on the mass $m_\pi$ or $m_q$ of internal loop
in the light-by-light scattering amplitude
were found to be as follows:
\begin{eqnarray}
 a_\mu (a)(xm_\pi , M_\rho) &\sim&
  2.81 \times 10^{-2} \times x^{-2}
  \left ( {\alpha \over \pi} \right )^3
 , \nonumber \\
 a_\mu (b)(xm_\pi , M_\rho) &\sim&
  -9.57 \times 10^{-2} \times x^{-2}
  \left ( {\alpha \over \pi} \right )^3
  ,   \nonumber \\
 a_\mu (c)(xm_q, M_\rho) &\sim&
 1.14\times 10^{-4}\times x^{-3.7}
  \left ( {\alpha \over \pi} \right )^3
  ,
 \label{asy1}
 \end{eqnarray}
\noindent
for $x > 3  $, where $x$ is a scale factor of pion mass
in $a_\mu (a)$ and $a_\mu (b)$, or of quark masses
(0.3, 0.3, 0.5, and 1.5 GeV for $u, d, s,$ and $c$,
respectively) in $a_\mu (c)$.
  The first result in (\ref{asy1}) shows
that the contribution of pion loop momenta
drops off as $x^{-2}$ as $x$ increases.
 For instance,
the contribution of pion momenta higher than 800 MeV
accounts for only 7 percent
of (\ref{sQED}).
 From these results, we  conclude that
the hadronic light-by-light scattering amplitude,
even when it is inserted
in Feynman graphs for muon anomaly,
can be described reasonably well by the graphs we have studied.

 We find the behavior  $a_\mu(c)\sim x^{-4}$ quite encouraging.
 The fact that Fig. 2(c) contributes at all,
 as is seen from (\ref{qloop}),
 implies that energy scale
of ${\cal O}$(1 GeV) is important.
 On the other hand,
the steep $x$ dependence of $a_\mu (c)$
found in (\ref{asy1}) is
consistent with the fact that only the physical degree of freedom
(mainly pions) is important at low energies\cite{stan}.

 We have also studied
the dependence of $a_\mu$ on the vector meson mass:
\begin{eqnarray}
 a_\mu(a)(m_\pi, M) &\sim&  [ -0.035~6
  + 0.23 \left ( {m_\mu \over M } \right )]
 \left ( {\alpha \over \pi} \right )^3
  ,  \nonumber \\
 a_\mu(b)(m_\pi, M) &\sim&  [-0.069~3
  + 0.31 \left ( {m_\mu \over M } \right )]
 \left ( {\alpha \over \pi} \right )^3
  ,   \nonumber \\
 a_\mu(c)(m_\pi, M) &\sim&  [+0.044~ 0
  - 0.43 \left ( {m_\mu \over M } \right )]
 \left ( {\alpha \over \pi} \right )^3
\label{asymp1}
\end{eqnarray}
for $M > M_\rho$.
 This shows that these functions decrease
very slowly for large $M$.
 Such an $M^{-1}$ (instead of $M^{-2}$) behavior
seems to cast some doubt
on the effectiveness of our approach
since it implies that an appreciable contribution to $a_\mu$ comes
from photon momenta far off mass-shell.
 Actually, this is entirely compatible
with the dominance of low energy states
in the light-by-light scattering amplitude.
 The naively expected $M^{-2}$ behavior cannot be justified unless
the $M_\rho \rightarrow \infty$ limit
and subintegrations in the Feynman diagram commute.
 As it is clear from power counting, this is not the case.
 Also, the results (\ref{asymp1})
show that there are strong cancellations between
the first and second terms.
 In particular the cancellation in $a_\mu (a)$
for $M=M_\rho$ is almost complete.

(e) Error estimates

 As stated above, three diagrams shown
in Fig. 2 seem indeed to dominate
the light-by-light scattering amplitude in the Feynman
graphs for the muon anomaly.
The contributions of kaon loop to $a_\mu (a)$,
$\eta$ pole to $a_\mu (b)$,
and heavier quarks to $a_\mu (c)$ can be readily included.
 We expect the error coming from
further additional diagrams,
as well as from the double counting possibility,
to be less than the error caused
by the approximate treatment of photon
propagators in the HLS approach.
 By using HLS, we have not taken account,
for example, of the continuum states above the vector mesons\cite{foot}.
 As long as we can restrict ourselves to pseudoscalar and quark loop
diagrams, however, it is hard to imagine that these continuum states
are relevant. Indeed the pion form factor is saturated with
$\rho$ meson.

 Taking these considerations into account,
we estimate that the model-dependent errors from
the terms of (\ref{qloop}) are well within
$20 \%$ of the $M_\rho$-dependent second term.
Including the contributions
of the kaon loop coupled
to $\phi$ in $a_\mu (a)$,
the $\eta$ pole to $a_\mu (b)$,
the strange and charm quarks to $a_\mu (c)$,
we are thus lead to
\begin{eqnarray}
  a_\mu(a)
 &=& - 0.003~6~(64) \left( \frac{\alpha}{\pi} \right)^3 ,
   \nonumber \\
  a_\mu(b)
 &=&  -0.032~8(66)
 \left( \frac{\alpha}{\pi} \right)^3 ,   \nonumber \\
 a_\mu(c) &=&
   ~~0.007~7(88) \left( \frac{\alpha}{\pi} \right)^3  .
 \label{res}
\end{eqnarray}

(f) Summary:

 Using chiral symmetry, $1/N_c$ expansion,
and hidden local symmetry as guides,
we have found that the hadronic light-by-light
scattering amplitude can be represented reasonably well
 as the sum of diagrams containing a
charged pion loop, diagrams with a $\pi^0$ pole,
and diagrams with a quark loop
at energies below 1 GeV.
 Based on this observation we have computed
the hadronic light-by-light scattering correction to the
muon $g-2$  due to three diagrams of Fig. 2.
 The integration over the photon momenta
receives considerable contribution from the region where
the photons are far off shell.
 Estimating that these high mass contributions
should be well
within 20\% of the vector meson contribution,
we have obtained the result (\ref{res}) which leads to
 \begin{equation}
 a_\mu({\rm light\mbox{-}by\mbox{-}light})
= -36(16)\times 10^{-11} .
 \end{equation}
 This is within the error expected
 in the upcoming experiment.
 Based on our analysis,
and in view of the progress
in the measurement of $R$ \cite{worstell},
we are quite hopeful
that the next round of experiments
will indeed verify the weak interaction
correction to the muon anomaly.

\vspace{0.2cm}
\centerline{Acknowledgments }

 T. K. thanks S. Tanabashi and H. Kawai for useful discussions.
He thanks the hospitality of the National Laboratory
for High Energy Physics (KEK), Japan,
where he was visiting on sabbatical leave from Cornell University.
 His work is supported in part by the U. S. National Science Foundation.
 One of the authors ( M. H. ) acknowledges support
in part from Toyota Physical and Chemical Research Institute.
%
%

%
%
%
%
\begin{center}
 FIGURE CAPTIONS
\end{center}
\vspace{0.2cm}
 Fig. 1.
 Hadronic light-by-light scattering contribution
 (shown by the shaded blob) to the muon anomaly.
 Solid line and dashed line represent muon and photon, respectively.

\vspace{0.5cm}
\noindent
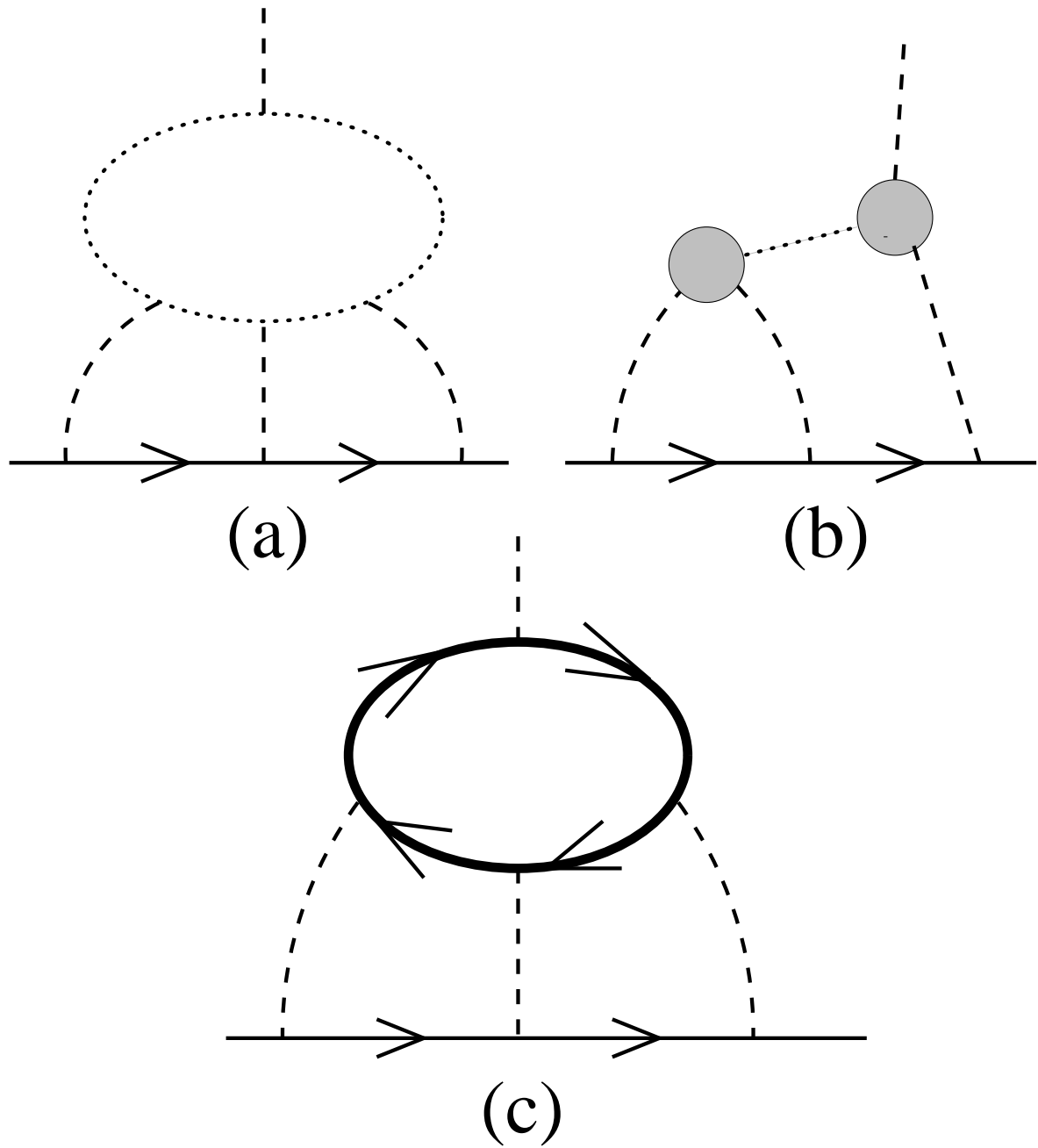
 Fig. 2. Diagrams which
 dominate the hadronic light-by-light effect on $ a_\mu $ at low energies.
 (a) Charged pseudoscalar diagram
 in which the dotted line corresponds to $ \pi^\pm $, etc.
 (b) One of the $ \pi^0 $ pole graphs,
 in which the dotted line corresponds to $ \pi^0 $
 and the blob represents the $ \pi \gamma \gamma $ vertex.
 (c) Quark loop contribution,
 where quark is denoted by bold line.

%
%
%


\begin{thebibliography}{99}
\bibitem{foot}
 We also expect, as in the ENJL model,
that a $q^2$-dependent function $M_V^2 (q^2)$,
where $q$ is the $\rho$ momentum,
appears in place of the constant $\rho$ meson mass.
 As is seen from the explicit form of $M_V^2 (q^2 )$
given in \cite{deRafael},
the $q^2$-dependence in the relevant range
($ \leq M_\rho$) is relatively mild
so that the correction to this approximation seems to be small.
\end{thebibliography}
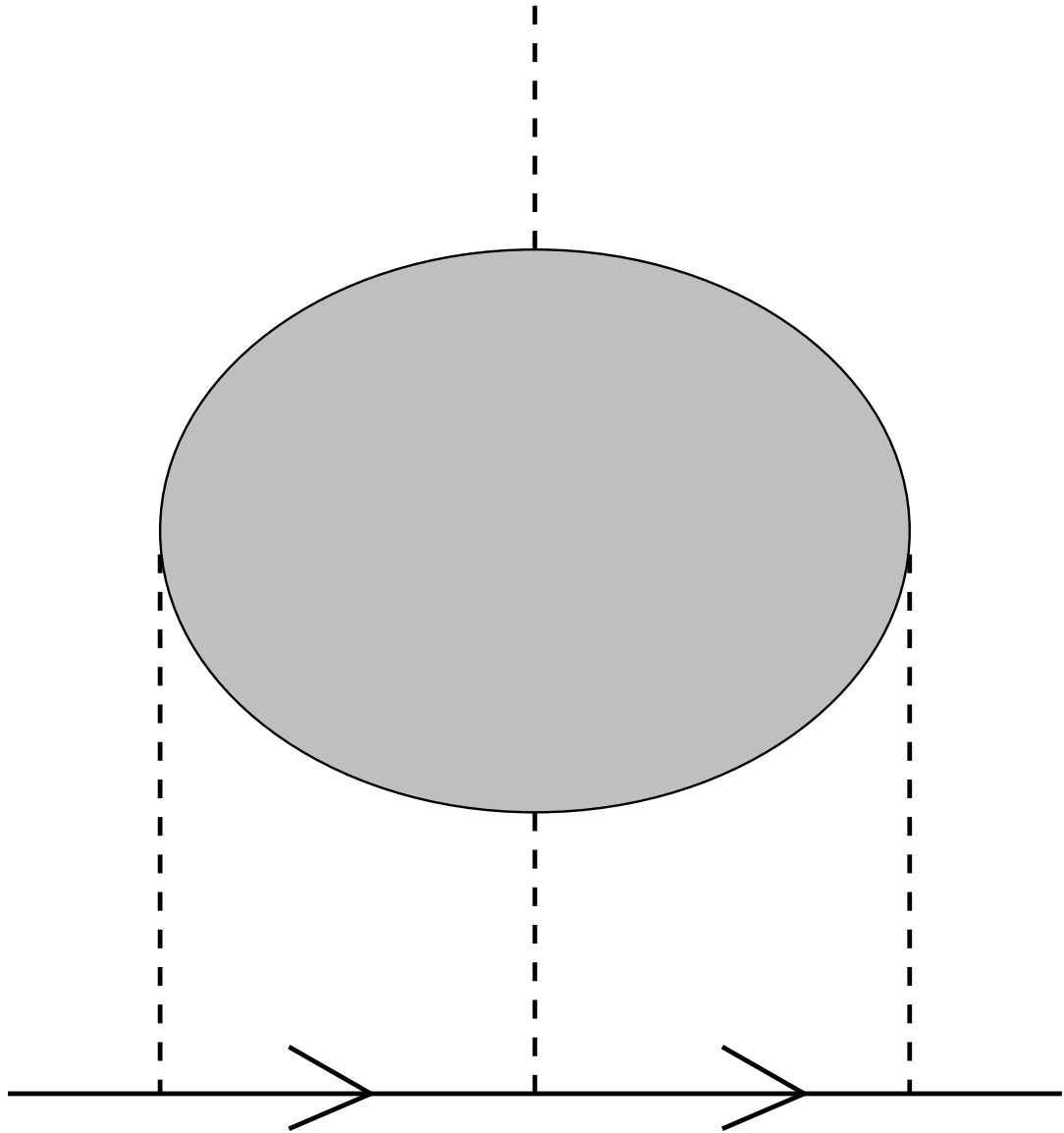
\end{document}